\documentclass[letterpaper, 10 pt, conference]{ieeeconf}  

\usepackage{graphicx}
\graphicspath{ {./} }
\usepackage{booktabs}
\usepackage{color}
\usepackage{float}

\title{\LARGE \bf
Understanding the COVID19 Outbreak: \\A Comparative Data Analytics and Study
}

\author{Anis Koubaa$^{1,2}$\\
{$^{1}$Robotics and Internet-of-Things Lab (RIOTU), Prince Sultan University, Riyadh, Saudi Arabia.} \\
{$^{2}$CISTER, INESC-TEC, ISEP, Polytechnic Institute of Porto, Portugal.}\\   
akoubaa@psu.edu.sa
}

\begin{document}
\maketitle
\thispagestyle{empty}
\pagestyle{empty}

\begin{abstract}
The Coronavirus, also known as the COVID-19 virus, has emerged in Wuhan China since late November 2019. Since that time, it has been spreading at large-scale until today all around the world. It is currently recognized as the world's most viral and severe epidemic spread in the last twenty years, as compared to Ebola 2014, MERS 2012, and SARS 2003. Despite being still in the middle of the outbreak, there is an urgent need to understand the impact of COVID-19. The objective is to clarify how it was spread so fast in a short time worldwide in unprecedented fashion. This paper represents a first initiative to achieve this goal, and it provides a comprehensive analytical study about the Coronavirus. The contribution of this paper consists in providing descriptive and predictive models that give insights into COVID-19 impact through the analysis of extensive data updated daily for the outbreak in all countries. We aim at answering several open questions: How does COVID-19 spread around the world? What is its impact in terms of confirmed and death cases at the continent, region, and country levels? How does its severity compare with other epidemic outbreaks, including Ebola 2014, MERS 2012, and SARS 2003? Is there a correlation between the number of confirmed cases and death cases? We present a comprehensive analytics visualization to address the questions mentioned above. To the best of our knowledge, this is the first systematic analytical papers that pave the way towards a better understanding of COVID-19. The analytical dashboards and collected data of this study are available online \cite{riotu-covid19}. 
\end{abstract}


\section{Introduction}
\label{sec:introduction}

The Coronavirus (COVID-19) outbreak nowadays represents the most critical event worldwide. It has been declared by the World Health Organization (WHO) as a Global Public Health Emergency by the end of January 2020, and then as a global pandemic in March 2020. The impressive fast spread of the virus is unprecedented and has exceeded all expectations. The containment of the virus is increasingly challenging as almost all countries in the world become infected. 

The virus has begun on from Wuhan district in China, where the first confirmed case was reported to have happened on November 17, 2020 \cite{chinacdc}. Initially, the confirmed cases in China were continually increasing. On January 31, the total infections reached a bit less than 10000 confirmed cases, with 214 recovered and 213 reported deaths (2\% death rate, and similar for recovery). Although the Chinese authorities have taken incremental and prompt preventive measures to avoid the exponential outbreak, the virus continued to spread not only within Chinese borders but also worldwide. The virus was transmitted through travelers around the world. One of the most dangerous aspects of the Coronavirus is that it has an incubation period of 2-14 days, during which the patient transmits the virus without having any symptoms. All these circumstances have favored the exponential growth of the infection leading to a world health emergency crisis. As a consequence, after only two months from the official declaration of COVID-19 as Global Public Health Emergency, and despite the numerous exceptional preventive measures that every country has taken to avoid the outbreak, the virus has contaminated almost all the world countries.

\begin{figure*}[ht]
	\centering
	\includegraphics[width=17cm]{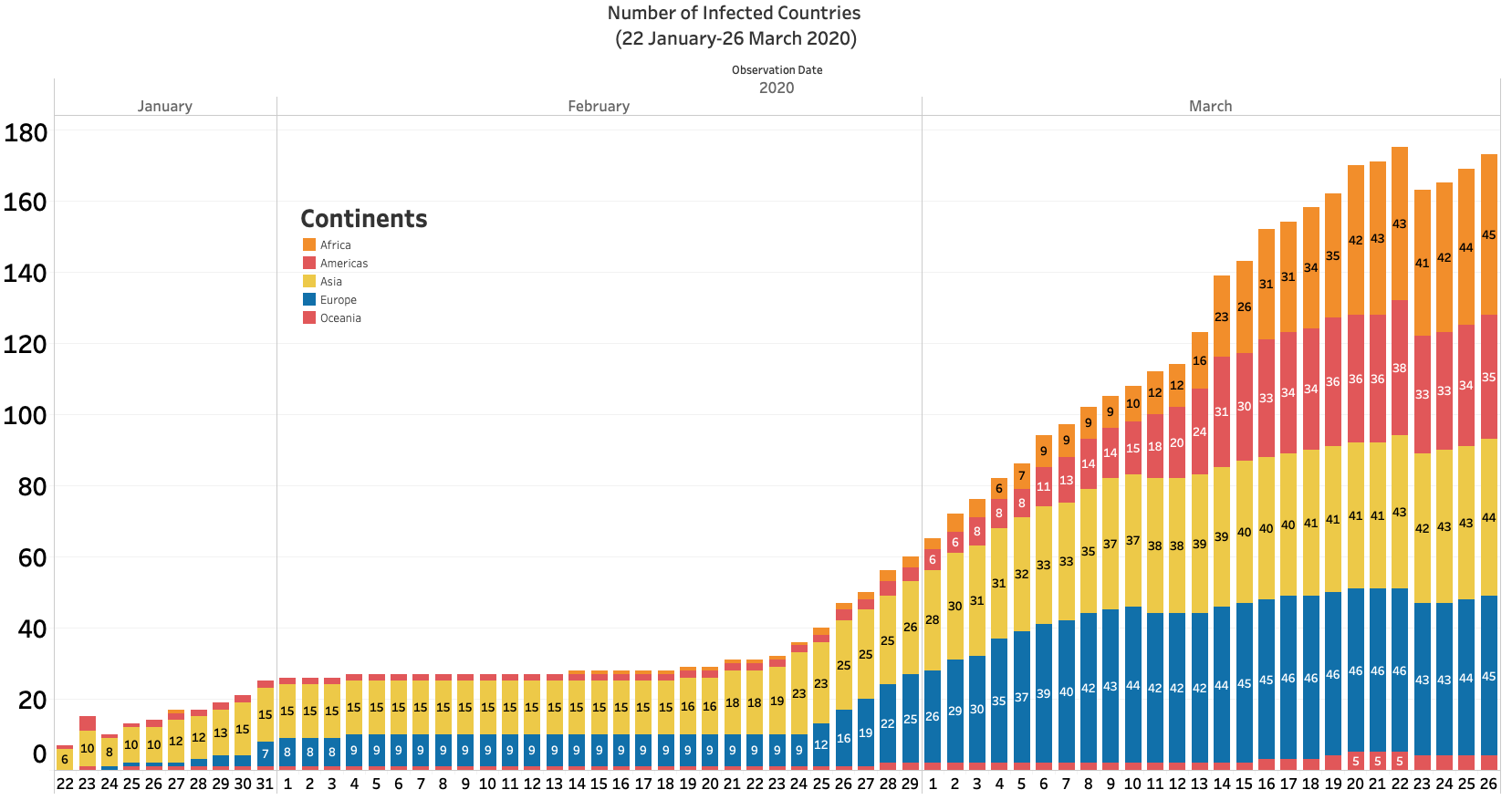}
	\caption{Evolution of the number of infected countries over time from 22 Jan until 26 March 2020}
	\label{fig:number-of-infected-countries}
\end{figure*}

Figure 1 illustrates the evolution of the number of countries that were affected by the Coronavirus outbreak from January 22 until March 26, 2020, based on the daily report data provided by Johns Hopkins University repository \cite{JohnsHopkinsNovelCoronavirusDataSet}.It is observed from the figure that the outbreak pick growth started towards the end of February, which is almost four weeks since the disease was declared as a Global Public Health Emergency. Besides Asia, the first continent to be severely affected, the outbreak has been more generalizing to other continents during March 2020, putting first Europe into a crisis, followed by the Americas, and finally the African countries. At the time of writing this paper, a total of 173 countries are reported to have confirmed cases with different gravity, while only 60 countries had confirmed cases at the end of February 2020, and only 25 countries at the end of January 2020. This means that the increase rate was between 2.4 to 2.9 each month. 
Almost all countries worldwide are currently infected, but the impact of the COVID-19 virus has widely varied between the continents, regions, and countries. This represents the motivation of this data analytics study. Our objective is to unveil the secrets of the COVID-19 virus and understand its evolution in the world. We aim to know the distribution of confirmed and death cases across the continents, regions, and countries and the correlation between them. Furthermore, we compare the impact of the COVID-19 virus with the previous and recent epidemic diseases that have emerged in the last 20 years, namely Ebola 2014, Middle East Respiratory Syndrome (MERS 2012), and Severe Acute Respiratory Syndrome (SARS 2003) outbreaks. In particular, we address the following research questions:  How does  COVID-19  spread around the world?  What is its impact in terms of confirmed and death cases at the continent, region,  and country levels?  How does its severity compare with other epidemic outbreaks, including Ebola 2014, MERS 2012, and SARS 2003? Is there a correlation between the number of confirmed cases and death cases?

The rest of the paper is organized as follows. 

\section{Related Works}
Since its spread, there have been several initiatives to investigate the impact of COVID-19 from the scientific community. 

In \cite{Li2020}, the authors have proposed to analyze the use of social media to exchange information about the Coronavirus. They proposed to identify situational information to investigate the propagation of COVID-19 related information in social media. They used natural language processing techniques to classify COVID-19 information into several types of situational information. 

In \cite{Joseph2020}, the authors develop a predictive model to forecast the propagation of COVID-19 in Wuhan and its impact on public health by considering the social preventive measures. Some other researchers, like in \cite{Cereda2020}, the authors have analyzed the COVID-19 outbreak during its early phases in Italy.  They provided estimates of the reproduction number and serial intervals. In \cite{singh2020agestructured}, the authors investigated the impact of preventive measures, such as social distancing, lockdown, in the containment of the virus outbreak. They developed prediction models that forecast how these measures can reduce the mortality impact of aged people. 
The authors of \cite{chen2020visual} addressed the question about how the virus has spread from the epicenter of Wuhan city to the whole world. They have also analyzed the impact of preventive measures such as quarantine and city closure in mitigating the adverse impact of the spread. The authors have demonstrated visual graphs and developed a mathematical model of the disease transmission pattern. 
In \cite{vattay2020predicting}, the author has analyzed the virus outbreak in Italy based on early data collected to predict the outcome of the process. He argued that there is a strong correlation between the situation in Italy and that of Hubei Province.

Some researchers have attempted to use deep learning and artificial intelligence in the context of COVID-19. In \cite{wang2020covidnet}, the authors have proposed COVID-Net, which is a deep convolutional neural network for the detection of COVID-19 infection from chest radiography images open-source dataset. The dataset contains 5941 chest radiography images of 2839 patient cases. In \cite{gozes2020rapid}, the authors have developed an image processing technique for the detection, quantification, and tracking of the COVID-19 virus. They utilized deep neural network models for the classification of suspected COVID-19 thoracic CT features, using data from 157 patients from the USA and China. The classification area under the curve (AUC) of the study was found to be 0.996. 
In \cite{ghoshal2020estimating}, the authors investigated drop-weights based Bayesian Convolutional Neural Networks (BCNN) and its effect on improving the performance of the diagnostic of COVID-19 chest X-ray. They showed that the uncertainty in prediction is highly correlated with the accuracy of prediction.

\section{Data Analytics Methodology}
In this paper, we propose a detailed data analytics study about the COVID-19 virus to understand its impact. Besides, we compare its severity against Ebola 2014, MERS 2012, and SARS 2003.

To achieve this objective, we have collected data from authentic sources and widely accepted by the scientific community. In what follows, we present the datasets used in this study. 

\subsection{Datasets}
We searched for datasets that provide credible data about the COVID-19 outbreak. The 2019 Novel Coronavirus COVID-19 Data Repository provided by Johns Hopkins University \cite{JohnsHopkinsNovelCoronavirusDataSet} is the most comprehensive, up-to-date, and complete dataset that gives daily reports of the COVID-19 outbreak, in terms of confirmed cases, death cases, and recovered cases. Besides, Johns Hopkins University maintains an active dashboard that reports daily updates of the Coronavirus \cite{CSSECoronavirus}. Also, the same dataset is being extensively used by the data science community of Kaggle to develop several analytics notebooks and dashboard about COVID-19 \cite{KaggleCOVID19}.

Each row in the COVID-19 dataset contains the following relevant data: 
\begin{itemize}
	\item\textbf{ Observation date:} it represents the date when the corresponding data row was reported. 
	\item \textbf{Country:} the country from where the data emerged
	\item \textbf{Confirmed cases:} the number of COVID-19 confirmed cases
	\item \textbf{Death cases:} the number of COVID-19 death  cases
	\item \textbf{Recovered cases:} the number of COVID-19 recovered cases
\end{itemize}

In addition to this data, we have processed the dataset to add additional information related to: 
\begin{itemize}
	\item \textbf{Continent:} the continent of the country related to the collected data. We considered five continents, including Africa, the Americas (north and south), Asia, Europe, and Oceania (Australia and New Zealand). 
	\item \textbf{Region:} the region is a level between country and continent. We considered the following regions in our study: (Northern/Southern/Eastern/Western/Middle) Africa, (Northern/Southern/Eastern/Western) Europe, (Norther/Southern/South-Eastern/Eastern/Western) Asia, (Northern/South/Central) America, Arabic Gulf, Caribbean, Australia, and New Zealand, Melanesia, and Micronesia. 
\end{itemize}

The mapping between countries and their corresponding regions and continents was performed using the following CSV file \footnote{https://www.kaggle.com/statchaitya/country-to-continent}.

We have also collected datasets for the other epidemic diseases, namely:

\begin{itemize}
	\item SARS 2003 Outbreak Complete Dataset \cite{sars2003dataset}
	\item MERS Outbreak Dataset 2012-2019 \cite{mers2012dataset}
	\item Ebola 2014-2016 Outbreak Complete Dataset \cite{ebola2014dataset}
\end{itemize}

All the dataset provides time-series information about confirmed and death cases per country per day during the observation period, except, the MERS dataset that provides only the final statistics of the disease for the confirmed cases (no death cases reports). We could not find any credible data source for the time series evolution of MERS, neither the death cases. 

We have processed these datasets to clean the data and also add the mapping of the countries to their region and continent to develop region-level and continent-level statistics. Also, we have created an all-in-one dataset with all data combined for comparative purposes.

\subsection{Methodology}
In this work, we have used Tableau Professional software to analyze the collected data and develop visualization dashboards about the Coronavirus disease.  
Our methodology consists in creating descriptive models of the Coronavirus outbreak using statistical charts to understand the nature of the spread and its impact. We develop our analysis at three levels, namely, at the country-level, at region-level, and continent-level. Each level provides different granularities towards understanding the distribution of the disease around the world. The descriptive model provides different types of statistical charts, including bar charts, geographic maps, heat maps, box plot, and packed bubbles, to represent different features of the COVID-19 outbreak. 
We also develop some predictive models using linear and polynomial regressions to predict the evolution of the outbreak, given the historical data. 

In this study, we also compare COVID-19 with the other three most critical world epidemic outbreaks, namely Ebola 2014, MERS 2012, and SARS 2003. We visualize the difference in terms of the impact of these diseases in terms of confirmed and death cases, analyze the characteristic of each disease. 

The lessons learned in this data analytics study serves as a ground for data science for further investigation of the COVID-19 epidemic outbreak.

\section{Results}
In this section, we will present the results of this data analytics study. The dashboards of this study are also available online \cite{riotu-covid19}.

\subsection{How does COVID-19 evolve?}

\begin{figure*}[ht]
	\centering
	\includegraphics[width=17cm]{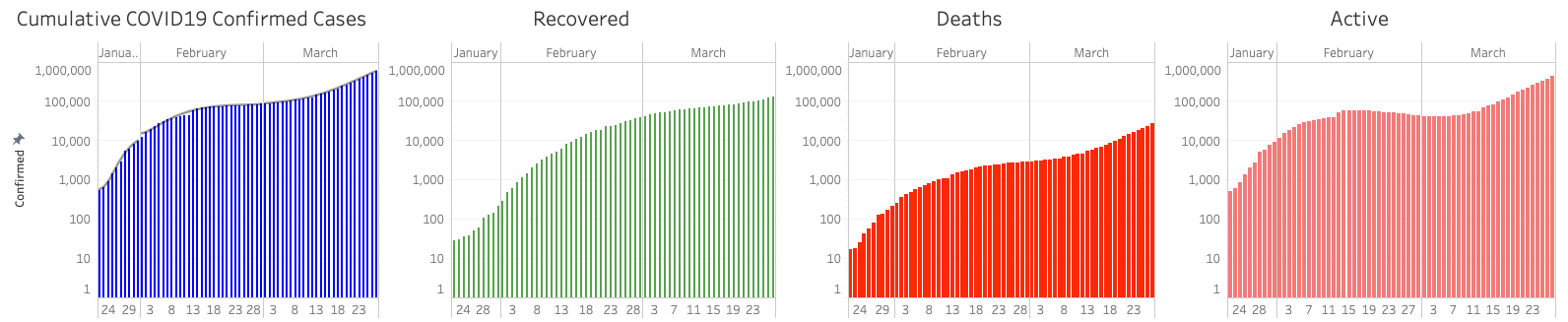}
	\caption{COVID-19 Dashboard 1: The four figures in the first row presents the cumulative confirmed, recovered, death and active cases from January 22, 2020 to March 27, 2020.}
	\label{fig:dashboard-cumulative-cases}
\end{figure*}

Figure \ref{fig:dashboard-cumulative-cases} depicts the evolution of the COVID-19 outbreak in the logarithmic scale during the period from January 22, 2020, to March 27, 2020, i.e., two months period. Let us consider January as the reference month. 

We performed a linear regression analysis on the different curves shown in Figure \ref{fig:dashboard-cumulative-cases}, and we determined the confirmed/recovered/death/active rates during the observation periods. These rates are shown in Table \ref{table:growth-rate}. The rates are the slope of the regression lines. They are shown as the first parameter between parenthesis in the table below. 

\begin{table}[]
	\caption{Linear regression parameters of cases growth rates at every month corresponds to the growth rate of every cumulative case}
	\label{table:growth-rate}
	\begin{tabular}{|l|l|l|l|}
		\hline
		\textbf{Cumulative Cases/Day} & \textbf{January 2020} & \textbf{February 2020} & \textbf{March 2020} \\ \hline
		\textbf{Confirmed Case Rate} & 1071 & 2730 & 16679 \\ \hline
		\textbf{Death Cases} & 22 & 103 & 792 \\ \hline
		\textbf{Recovered Cases} & 19 & 1373 & 2906 \\ \hline
		\textbf{Active Cases} & 1029 & 1252 & 12890 \\ \hline
	\end{tabular}
\end{table}

By observing the trend lines of linear regression models on the different curve at each month, we conclude the following observation:

\begin{itemize}
	\item The global cumulative confirmed case rates are 2.54 times higher in February 2020, and 15.67 times higher in March 2020, as compared to the confirmed case rate of January 2020.  
	\item The global cumulative death case rates are 4.68 times higher in February 2020, and 36 times higher in March 2020, as compared to the death case rate of January 2020. This indicates the severe and increasing fatality impact of the COVID-19. Death rates in February were stationary at an average of 92 deaths per day. However, in March 2020, the death rates had a factor of 108. 
	\item The global cumulative recovered case rates are 72.26 times higher in February 2020, and 152 times higher in March 2020, as compared to the recovered case rate of January 2020. This indicates the severe and increasing fatality impact of the COVID-19. This is a good indication that, with the increase in the number of cases, there is a better understanding of the disease and containment. 
\end{itemize}

This is also confirmed by the ratio between the recovered rate and the death rate. 
Comparing the trend death and the trend of recovered cases, it can be observed that the ratio of recovered to death rates, was 0.86 (19/22) in January 2020, meaning that the death rate was a bit higher in January than the recovered rate. However, in February 2020, the ratio of recovered to death rates increases to 13.33 (1373/103) and reaches 3.66 (2906/792) in March 2020. Thus, the general trend is that the disease is being more controlled in terms of fatality rates due to increasing emergency procedures that the different countries have implemented. 

The results presented above a coarse grain in the sense that they related a global assessment of the evolution. However, the evolution of the COVID-19 infection depends much on the countries, the region, and the continent. It is, therefore, important to assess the evolution at these levels to get a better understanding of it. Figure \ref{fig:country-region-cases} presents the cumulative confirmed/recovered/death/active cases reported as of March 27, 2020 for the top-10 countries, then their regions and continents. 

We observe that in the top-10 countries, there are six countries are from Europe (i.e., Italy, Spain, Germany, France, United Kingdom, and Switzerland), three from Asia (China, Iran and South Korea), and the United States of America, which recently become top-1 in terms of the number of infections. Nonetheless, the highest death rate is in Italy, with more than 9000 death cases reported, because Italy has been severely affected by the virus well before the USA, since the end of February. However, the USA is currently having a death rate of 39 deaths per day, whereas it has 0 deaths in February 2020.

\begin{figure}[ht]
	\centering
	\includegraphics[width=8cm]{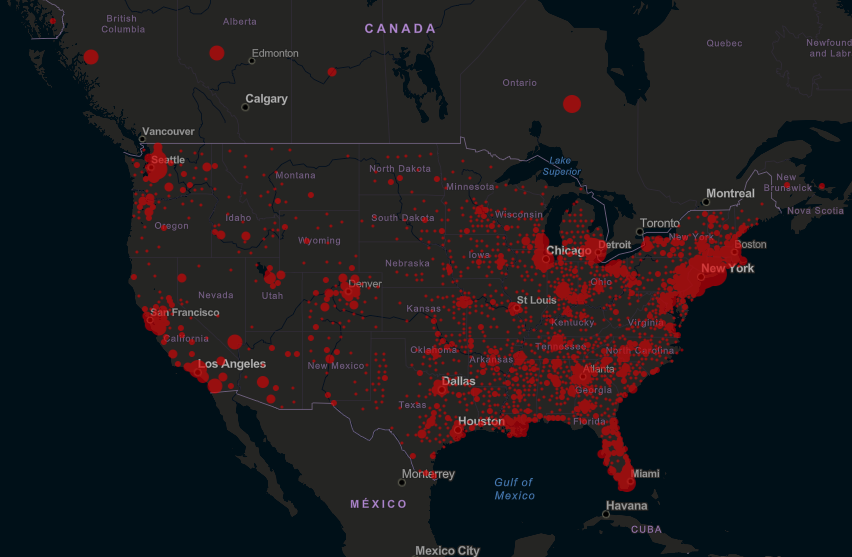}
	\caption{COVID-19 Outbreak in the USA as of March 27, 2020. It is clear that the Eastern side, and mainly New York are the most affected, considering it is closer to Europe.}
	\label{fig:usa-27-march-2020}
\end{figure}

\begin{figure*}[ht]
	\centering
	\includegraphics[width=17cm]{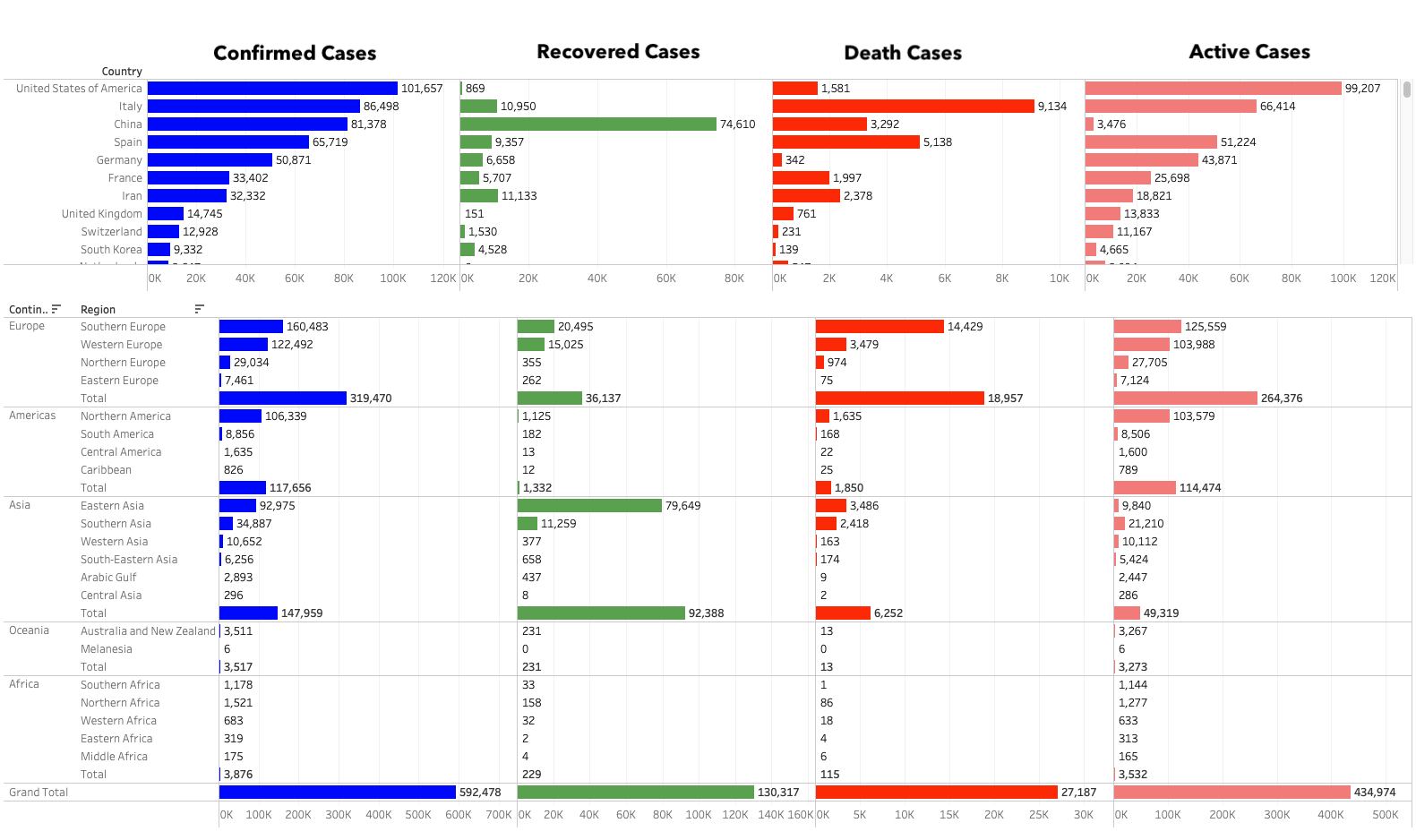}
	\caption{COVID-19 Dashboard 2 (as of March 27, 2020): The four figures in the first row presents the cumulative confirmed, recovered, death and active cases for the top-10 countries. In the second row, the same cases are presented at region and continent level}
	\label{fig:country-region-cases}
\end{figure*}

Based on the data collected there is a strongly believed that the COVID-19 \textit{takes almost one month} to transit from one continent to another in the direction from the East to the West since the pick was in China at the end of January, then it was in Italy (South of Europe) at the end of February, and it reached the USA at the end of March, where the pick of infection are in New York located at the Eastern side of the USA.  It can be observed in Figure \ref{fig:usa-27-march-2020} that the Eastern side of the USA, and mainly New York, are the most affected, considering that it is closer to Europe. If the trend is the same, it will be expected that the West side of the USA will reach its pick by the end of April 2020. 

In Figure \ref{fig:country-region-cases}, it can be observed that Europe has the most significant confirmed cases currently and the most severe fatality rates, where the maximum reached are the south of Europe with more than 14000 death cases, among which more than 9000 are located in Italy. Italy is currently having the third of fatalities in the whole world. 

It is also observed that Oceania and Africa are less affected by the virus as compared to other continents.

\begin{figure}[ht]
	\centering
	\includegraphics[width=9cm]{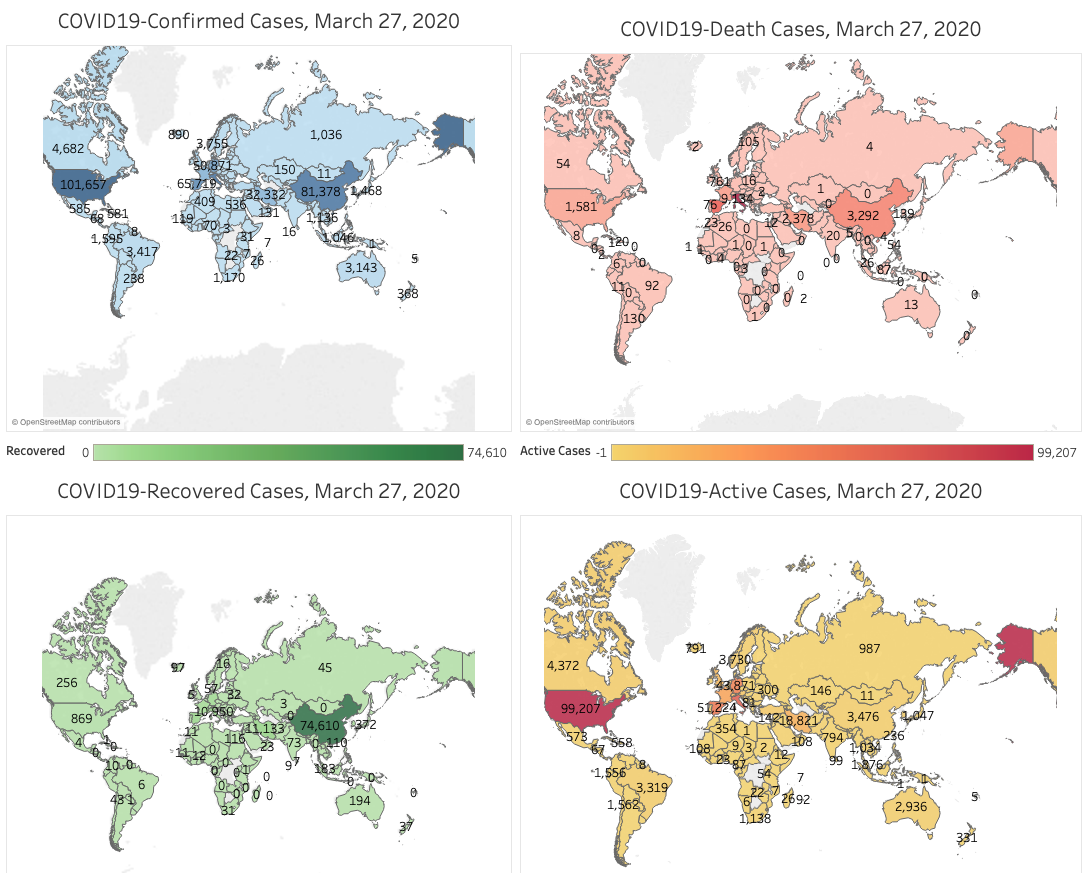}
	\caption{COVID-19 Outbreak Geographical Heatmap}
	\label{fig:covid19-map-spread}
\end{figure}

Finally, the distribution of the different cases is illustrated in the heatmap presented in Figure \ref{fig:covid19-map-spread}. Dark colors mean a high concentration of cases, and lighter colors mean smaller concentrations of cases. 

\subsection{How does COVID-19 compare to other epidemics?}

In the previous section, we have presented a comprehensive analysis of the COVID-19 virus, and we got a better understanding of how it was evolved and its impact on the country, region, and continent level. In this section, we address the question: How does COVID-19 compare to other epidemics?

Several other epidemics have emerged in the last 20 years, in particular, Severe acute respiratory syndrome (SARS2003) in 2003 in Hong Kong, the Middle East respiratory syndrome (MERS2012) in Saudi Arabia and the Middle East, and Ebola 2014 in Western African coast, namely Guinea, Liberia, and Sierra Leone. These three epidemics, in addition to COVID-19, are the most remarkable world diseases in the last 20 years, which we proposed to compare and analyze. 

\subsubsection{Comparative evolution over time}

\begin{figure*}[ht]
	\centering
	\includegraphics[width=17cm]{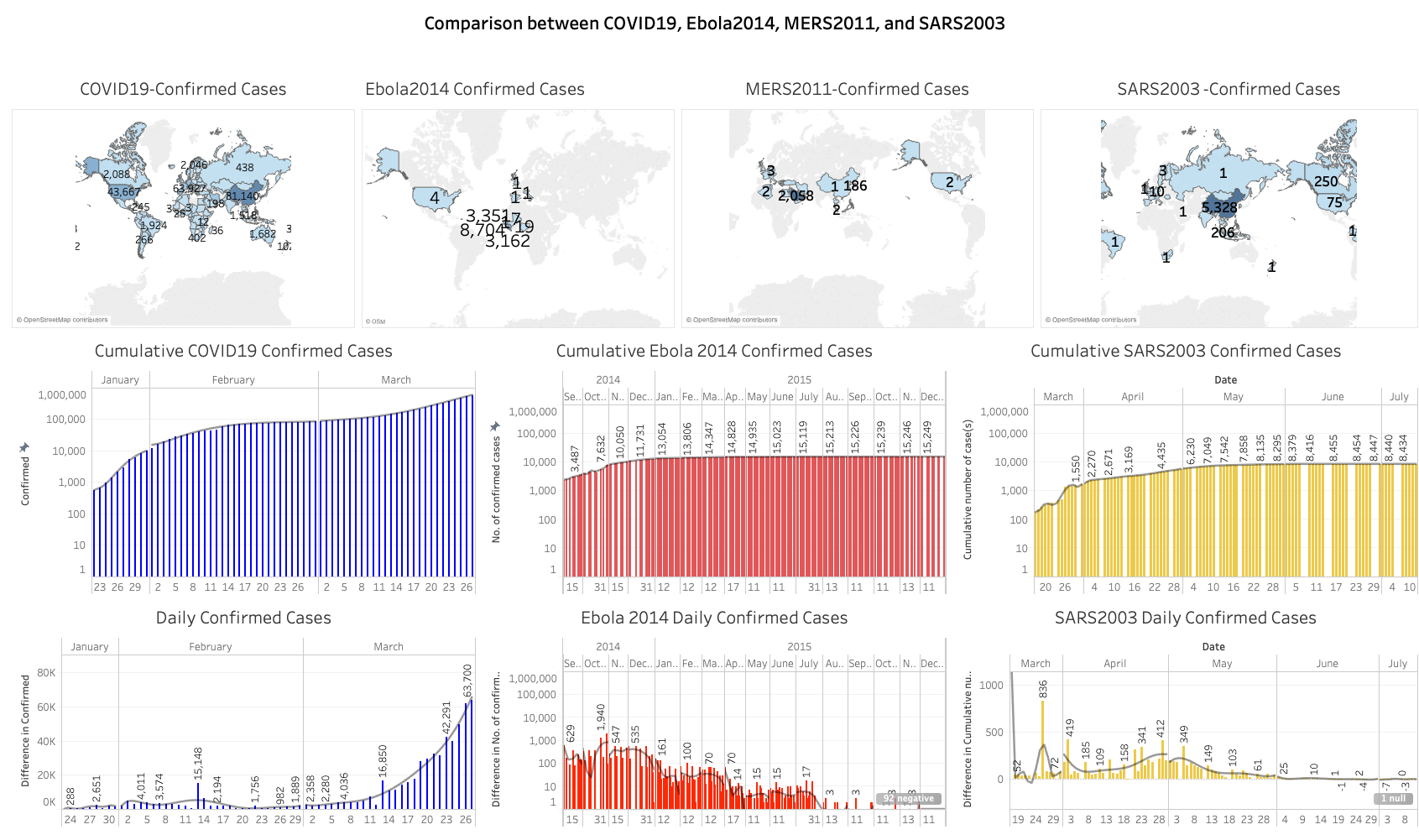}
	\caption{Comparative Dashboard between COVID-19, Ebola 2014, MERS 2011, and SARS 2003}
	\label{fig:dashboard-comparison-outbreak}
\end{figure*}

Figure \ref{fig:dashboard-comparison-outbreak} presents a dashboard that compares the four epidemic outbreaks. On the top, we observe the geographic heat map for the four diseases. It is visually apparent that COVID-19 is the largest outbreak to a considerable extent, followed by SAR 2003, then Ebola 2012, and finally, MERS 2001. In what concerns the number of infected countries, COVID-19 has reached 177 countries, then SARS 2003 reached 36 countries, then MERS 2012 has affected 27 countries, and Ebola 2014 was spread in 10 countries. The severity and acuteness of COVID-19 are unprecedented. 

We observe in the second row of Figure \ref{fig:dashboard-comparison-outbreak} that while Ebola 2014 has spanned over a significant period from August 2014 until March 2016, it reached only 15249 confirmed cases. Thus it had an infection rate minimal as compared to the two other diseases. Besides, most of the confirmed cases were concentrated in the Western coast of Africa, where the disease has emerged.  
On the other hand, SARS 2003 had a lifetime of five months from March 2003 until July 2003 and reached 8434 confirmed cases with an average of 1686 cases per month, and a pick rate in April 2003, one month after the outbreak. 

It can be observed that COVID-19 is considered as a more acute specie of SARS 2003, as they both share some common features, including: (1) both have started from China, (2) they belong to the same family of Coronavirus syndrome affecting the respiratory system, (3) they have the highest contamination rate as compared to other epidemics. 

Based on these observations, it seems that the COVID-19 containment will take a more extended period for its complete containment as compared to SARS 2003.

The third row of Figure \ref{fig:dashboard-comparison-outbreak} shows the daily confirmed cases for COVID-19, Ebola 2014, and SARS 2003. The trends of COVID-19 are exponential,  whereas the trends of Ebola 2014 and SARS 2003 are high at the start of disease then start to decrease after two months of the first confirmed cases. This shows that the behavior of COVID-19 is more aggressive as compared to the other epidemics.  

\subsubsection{Comparative Impact}
We address the question: how do the impacts of the epidemics compare to each other in terms of confirmed cases and death cases? 

\begin{figure*}[ht]
	\centering
	\includegraphics[width=17cm]{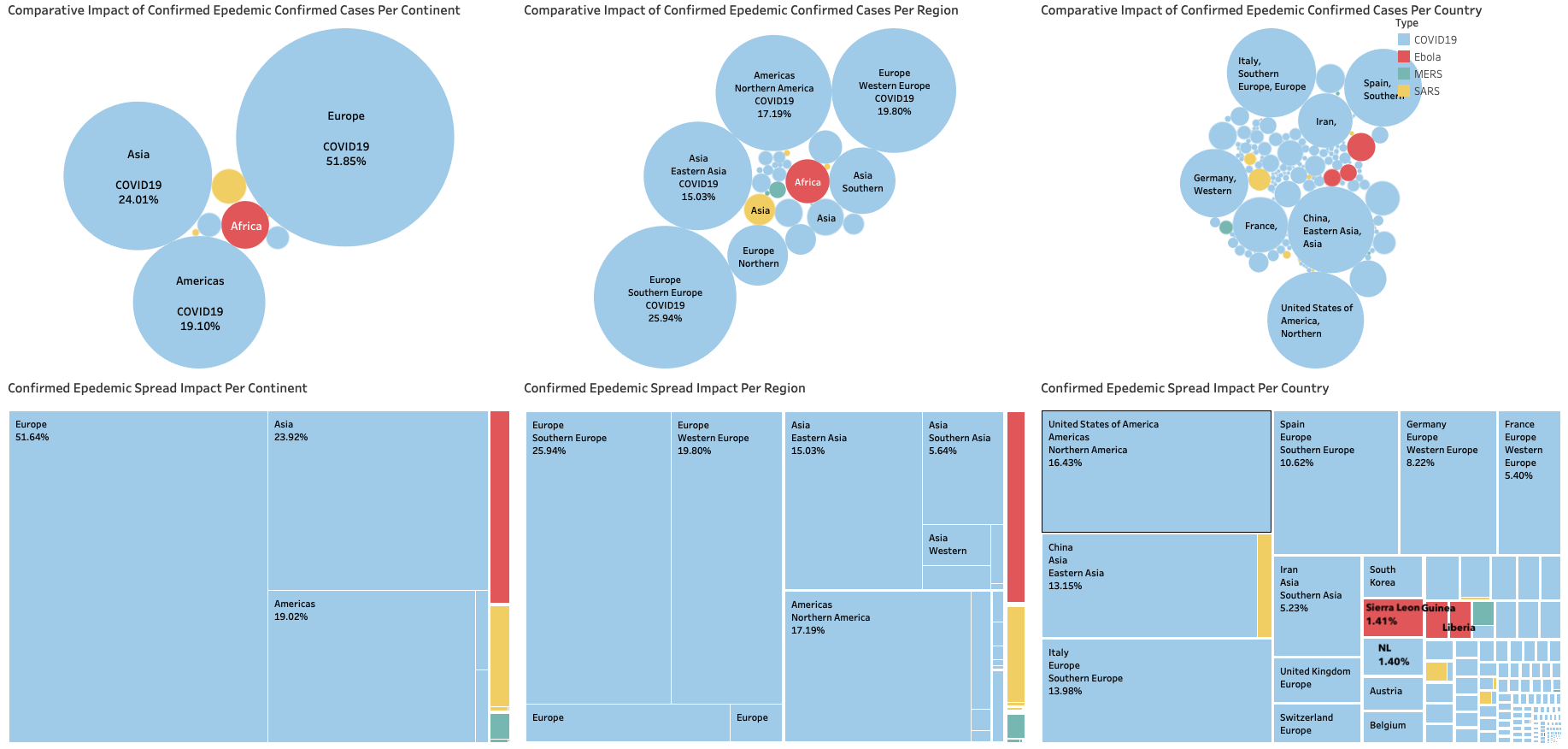}
	\caption{\textbf{Confirmed Cases: }Comparative Heatmap and Packed Bubble of the Impact of COVID-19, Ebola 2014, MERS 2011, and SARS 2003, at Continent-Level, Region-Level, and Country Level. (Note: COVID-19 as of March 27, 2020)}
	\label{fig:heatmap-confirmed-comparison-bubbles}
\end{figure*}

Figure \ref{fig:heatmap-confirmed-comparison-bubbles} shows the comparative impact with respect to the confirmed cases, and  Figure \ref{fig:heatmap-bubble-death-comparison} shows the comparative impact with respect to the death cases, at continent-level, region-level, and country-level. The blue color refers to the COVID-19; the red color refers to the Ebola 2014, and the yellow color refers to SARS 2003.  

Looking at the two figures, we can conclude that the COVID-19 is exceptionally more aggressive in terms of confirmed cases with more than 90\% of the share of the heatmap, where it is at around 80\% concerning the fatality impact. The remaining 10\% of confirmed cases and 20\% of death cases are shared between Ebola 2014 and SARS 2003. The results illustrate well the magnitude of the severity of COVID-19 as compared to the other diseases. 

At continent-level, Europe is the most affected with COVID-19 with 51.64\% of confirmed cases, then Asia with 23.92\%, and the Americas with 19.02\%. The impact of Ebola on Africa is only 2.46\%, and the impact of SARS 2003 on Asia is only 1.26 \%.

At region-level, the Southern Europe region has 25.94\% of confirmed cases, followed by Western Europe Asia 19.8\%, then Northern America 17.19\%, and Eastern Asia with 15\%. The impact of Ebola 2014 on Western Africa is valued to 2.46\%, and the impact of SARS Eastern Asia is only 1.26\%. 

At the country-level, The USA has the most significant share of confirmed cases (as of March 27, 2020) with 16.43\%, followed by Italy 13.98\%, then China 13.15\%. We can also observe that the number of confirmed cases of Ebola 2014 in Sierra Leon is similar to the COVID-19 spread in South Korea and countries in the West of Europe, namely, Netherlands, Belgium, and Austria.  

Regarding the death cases' impact, it is different from the confirmed cases. 

At continent-level, the highest death impact is in Europe with 56.26\%, then Asia with 18.56\% with COVID-19, which is of the same magnitude as the fatality of Ebola 2014 in Africa. 

Looking at Table \ref{table:age}, we can observe a strong correlation between the median age at a continent/region and the fatality rate. Europe is the oldest of all continents, with a median age of 42\% has the highest fatality rates, mainly in Southern and Western Europe.  

At region-level, we observed that the deadly impact of Ebola 2014 on Western Africa is the second most severe after the deadly impact of death in the South of Europe. 

At the country-level, the impact of COVID-19 is the highest in Italy, followed by the impact of Ebola 2014 in Sierra Leone. 

In what concerns SARS 2003, its fatality rate is much lower than Eolba 2014 and COVID-19 diseases.

\begin{table}[]
	\begin{tabular}{|l|l|}
		\hline
		\textbf{Continent} & \textbf{Median Age} \\ \hline
		\textbf{Europe} & 42 years \\ \hline
		\textbf{North America} & 35 years \\ \hline
		\textbf{Oceania} & 33 years \\ \hline
		\textbf{Asia} & 31 years \\ \hline
		\textbf{South America} & 31 years \\ \hline
		\textbf{Africa} & 18 years \\ \hline
	\end{tabular}
	\caption{Median Age per Continent}
	\label{table:age}
\end{table}

\begin{figure}[ht]
	\centering
	\includegraphics[width=8cm]{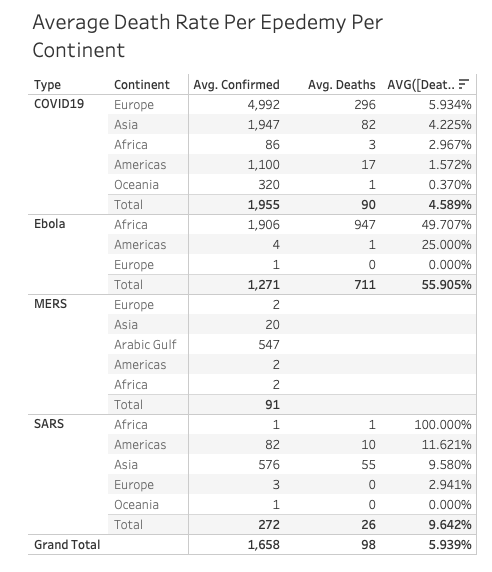}
	\caption{Average Death Cases Table (Note: COVID-19 as of March 27, 2020)}
	\label{table:average-deaths-table}
\end{figure}

\begin{figure*}[ht]
	\centering
	\includegraphics[width=17cm]{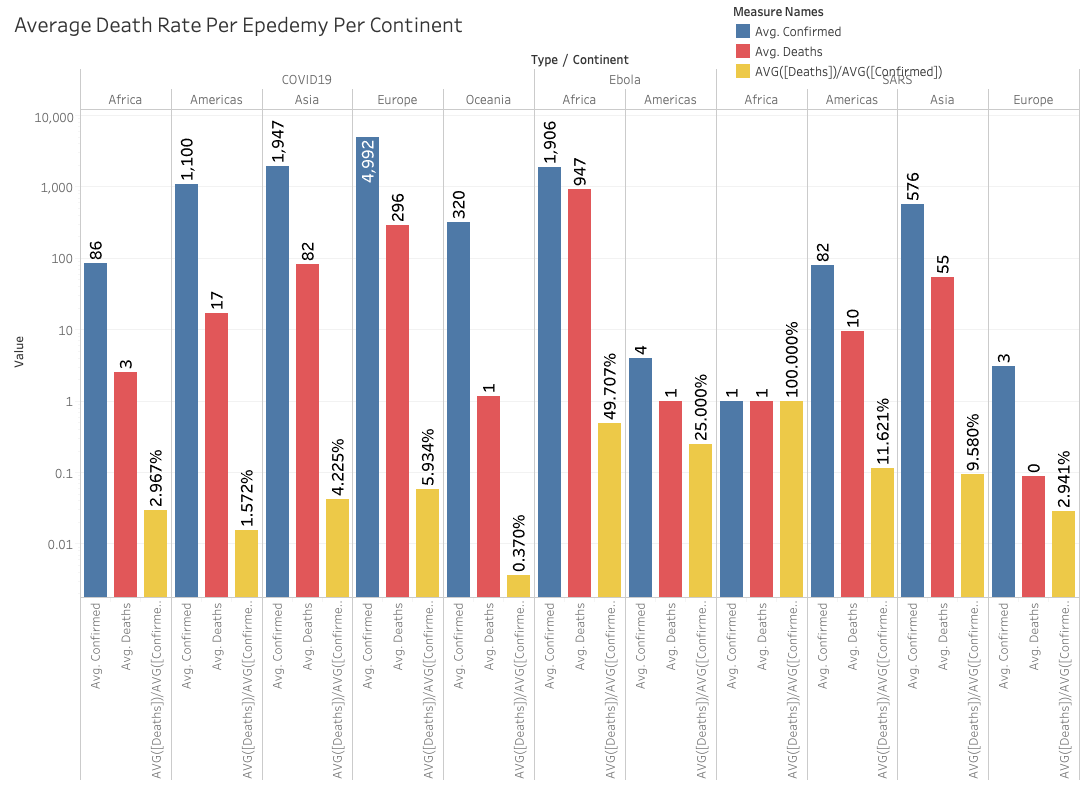}
	\caption{Average Death Cases Bar chart, logarithmic scale (Note: COVID-19 as of March 27, 2020)}
	\label{fig:average-deaths-chart}
\end{figure*}

\begin{figure*}[ht]
	\centering
	\includegraphics[width=17cm]{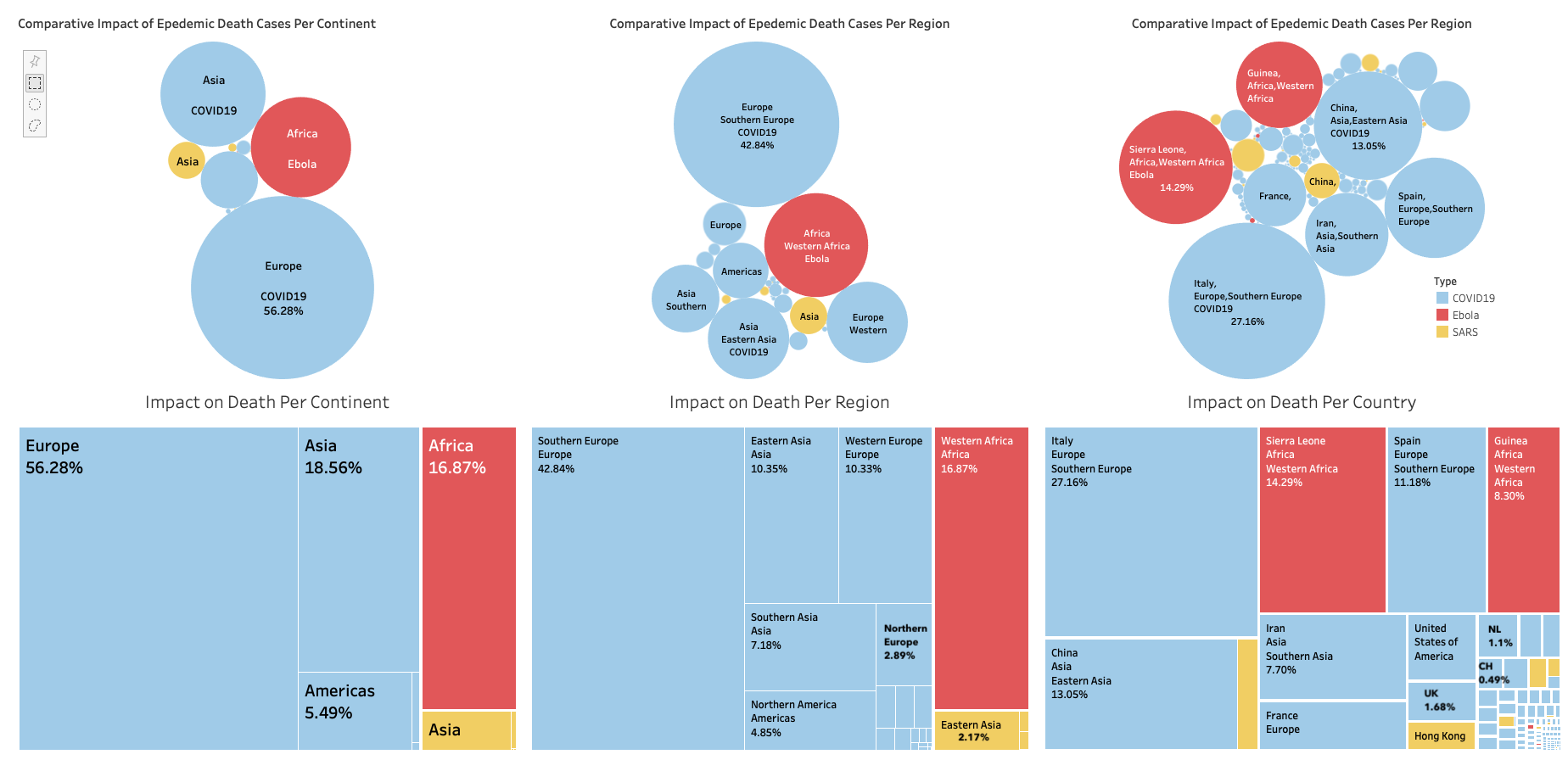}
	\caption{\textbf{Death Cases: }Comparative Heatmap and Packed Bubble of the Impact of COVID-19, Ebola 2014, MERS 2011, and SARS 2003, at Continent-Level, Region-Level, and Country Level. (Note: COVID-19 as of March 27, 2020)}
	\label{fig:heatmap-bubble-death-comparison}
\end{figure*}

Figure \ref{fig:average-deaths-chart} and Table \ref{table:average-deaths-table} present the average confirmed/death cases per continent for each of the epidemics per continent. The results confirm the heatmap, and packed bubbles presented above and provide the average distribution of death in each continent. 
The highest average of confirmed cases per continent is in Europe, with 4992 deaths of COVID-19, and the second-highest average confirmed cases per continent is 1906 deaths in Africa with Ebola 2014. MERS and SARS have the same average confirmed cases. 
Regarding the average deaths per continent, Ebola 2014 comes in the first place with 947 deaths per country in Africa on average, followed by COVID-19, with 296 deaths in each European country on average. SARS 2003 only affected 55 deaths per country in Asia on average. 

\section*{Acknowledgments}
This work is supported by the Robotics and Internet of Things Lab of Prince Sultan University.

\section*{Author Biography}
Anis Koubaa is a Professor in Computer Science in Prince Sultan University. He is the Director of the Research and Initiatives Center at Prince Sultan University, and the leader of the Robotics and Internet-of-Things Lab. He is also a senior research associate with CISTER Research Center at the Polytechnic Institute of Porto in Portugal. He received his PhD degree in 2004 from the University of Lorraine in France. His interest interest deals with deep learning, data science, Internet-of-Things, Unmanned Aerial Systems, and Mobile Robots. 

\bibliographystyle{ieeetr}
\bibliography{references, aniskoubaapublications}

\end{document}